\documentclass[sigconf, nonacm]{acmart}

\usepackage{makecell}

\usepackage{xcolor}
\definecolor{lightergray}{RGB}{240, 240, 240}

\usepackage{listings}
\newcommand{\per}{\textit{Against the Person}}
\newcommand{\aut}{\textit{Appeal to Authority}}
\newcommand{\pop}{\textit{Appeal to Popularity}}
\newcommand{\emo}{\textit{Appeal to Emotion}}
\newcommand{\gen}{\textit{Hasty Generalization}}
\newcommand{\cau}{\textit{Questionable Cause}}
\newcommand{\her}{\textit{Red Herring}}
\newcommand{\non}{\textit{None}}

\AtBeginDocument{
  \providecommand\BibTeX{{
    \normalfont B\kern-0.5em{\scshape i\kern-0.25em b}\kern-0.8em\TeX}}}

\begin{document}

\title[Evaluation of an LLM in Identifying Logical Fallacies]{Evaluation of an LLM in Identifying Logical Fallacies: A Call for Rigor When Adopting LLMs in HCI Research}

\author{Gionnieve Lim}
\email{gionnievelim@gmail.com}
\orcid{0000-0002-8399-1633}
\affiliation{
    \institution{Singapore University of Technology and Design}
    \country{Singapore}
}

\author{Simon T. Perrault}
\email{perrault.simon@gmail.com}
\orcid{0000-0002-3105-9350}
\affiliation{
    \institution{Singapore University of Technology and Design}
    \country{Singapore}
}

\renewcommand{\shortauthors}{Lim and Perrault}

\begin{abstract}
There is increasing interest in the adoption of LLMs in HCI research. However, LLMs may often be regarded as a panacea because of their powerful capabilities with an accompanying oversight on whether they are suitable for their intended tasks. We contend that LLMs should be adopted in a critical manner following rigorous evaluation. Accordingly, we present the evaluation of an LLM in identifying logical fallacies that will form part of a digital misinformation intervention. By comparing to a labeled dataset, we found that GPT-4 achieves an accuracy of 0.79, and for our intended use case that excludes invalid or unidentified instances, an accuracy of 0.90. This gives us the confidence to proceed with the application of the LLM while keeping in mind the areas where it still falls short. The paper describes our evaluation approach, results and reflections on the use of the LLM for our intended task.
\end{abstract}



\maketitle

\section{Background}

LLMs have been rather successfully employed in various fields and for various applications. Their low barrier to entry, where only a prompt is needed at the highest level, makes them an appealing technology for adoption. Due to their powerful natural language capabilities, LLMs have been used for a variety of tasks including question answering, machine translation, generation, prediction and classification. Focusing on classification tasks, the applications are diverse. LLMs can be used to classify textual~\cite{wang2023large}, tabular~\cite{hegselmann2023} and image~\cite{Pratt2023} data. And they are employed for various operations like data annotation~\cite{li2024comparative}, sentiment analysis~\cite{zhang2023sentiment} and information extraction~\cite{goel2023}. Leveraging on the classification capabilities of LLMs, we considered if they can be used to identify logical fallacies in texts, a function that would serve as part of a digital misinformation intervention.

While a simple test of prompting an LLM with various texts and asking it to identify logical fallacies led to \textit{seemingly correct} results, we could not be convinced that the LLM was truly capable of this task. Aligned with the principles of trustworthy AI, particularly on the transparency of the performance of the AI~\cite{Li2023}, a more rigorous evaluation was called for. To that end, we looked at verifying the suitability of the LLM for identifying logical fallacies by using a more rigorous approach of testing against a labeled dataset. The goal was to understand to what extent the LLM performs for the task and to obtain an accuracy metric that could be conveyed to users so that they are aware of the performance of the LLM and can adjust their expectations for the digital misinformation intervention.

\section{The Logical Fallacies}

To understand whether LLMs are suitable for the task of identifying logical fallacies, we conducted an evaluation of the classification performance of an LLM on a limited set of logical fallacies. Since there are over 300 logical fallacies~\cite{bennett2012logically}, we focused only on a reduced set which is more practical and replicable. We referred to work that classified, identified or used logical fallacies in online discussions~\cite{sahai2021breaking, Hidayat2020, nikolaidis2023experiments} and misinformation~\cite{Musi2022, Beisecker2024, Hruschka2023, Cook2018, bonial2022search, jin2022logical, Sourati2023, Lundy2023} literature and selected the most commonly mentioned fallacies (\per{}, \aut{}, \pop{}, \emo{}, \gen{}, \cau{}, and \her{}).
Table~\ref{tab:logfal} shows the definitions of these logical fallacies.

\begin{table*}[htb!]
  \caption{Common logical fallacies found in online discussions and misinformation.}
  \label{tab:logfal}
  \renewcommand{\arraystretch}{1.3}
  \begin{tabular}{cp{.55\linewidth}}
    \toprule
    \makecell{Logical Fallacy \textit{(in Latin)}} & Definition\\
    \midrule
    \makecell[t]{Against the Person\\\textit{(Argumentum Ad Hominem)}} & Attacking the person or some aspect of the person making the argument instead of addressing the argument directly.\\
     \makecell[t]{Appeal to Authority\\\textit{(Argumentum Ad Verecundiam)}} & Using an alleged authority who is not really an authority on the facts relevant to the argument as evidence.\\
    \makecell[t]{Appeal to Popularity\\\textit{(Argumentum Ad Populum)}} & Affirming that something is real or better because the majority in general or of a particular group thinks so.\\
    \makecell[t]{Appeal to Emotion\\\textit{(Argumentum Ad Passiones)}} & Manipulating the reader’s emotions in order to win the argument in place of a valid reason.\\
    \makecell[t]{Hasty Generalization\\\textit{(Secundum Quid)}} & Drawing a conclusion about all or many instances of a phenomenon on the basis of one or a few instances of that phenomenon.\\
    \makecell[t]{Questionable Cause\\\textit{(Non Causa Pro Causa)}} & Concluding that one thing caused another simply because they are regularly associated.\\
    \makecell[t]{Red Herring\\\textit{(Ignoratio Elenchi)}} & Attempting to divert the reader’s attention from the original argument by offering a different point.\\
    \bottomrule
  \end{tabular}
\end{table*}

\section{Technical Evaluation}

We used the LOGIC dataset by Jin et al.~\cite{jin2022logical} that contained 2,449 logical fallacy instances across 13 logical fallacy types to evaluate the LLM. To select only the relevant data, we removed data that (1) were duplicates, (2) did not pertain to our set of logical fallacies, (3) only defined or described logical fallacies, (4) contained Latin phrases, and (5) contained quizzing phrases. The last was due to the LOGIC dataset being crawled from student quiz websites. This resulted in a reduced dataset ($N=801$).

OpenAI's GPT-4 model\footnote{https://platform.openai.com/docs/models/gpt-4} was used as the LLM for identifying logical fallacies. We used the Chat Completions API with the \texttt{gpt-4} model, a temperature setting of \texttt{0}, and the role: \texttt{You are a critical thinker.} Few-shot prompting~\cite{Brown2020} was used for the logical fallacies classification task where three examples of each of the seven logical fallacies were provided. These instances ($n=21$) were taken from the dataset and excluded from the evaluation of the performance of the LLM. The breakdown of the full data used for evaluation ($N=780$) was thus as follows: 157 \per{}, 74 \aut{}, 133 \pop{}, 41 \emo{}, 165 \gen{}, 126 \cau{}, and 84 \her{}. The prompt used for identifying logical fallacies is shown in Appendix~\ref{prompt}.

\section{Results}

We present the multi-class classification performance metrics~\cite{grandini2020metrics} of the LLM in the identification of logical fallacies in Table~\ref{tab:perfmetrics}. When prompting, we allowed the LLM to answer \non{} if no logical fallacy was identified in a text. In rare instances where the LLM answered with a logical fallacy outside our considered set, we also post-processed the result to \non{}. As such, we further present the performance metrics on the subset ($N=685$) of the data where instances predicted or post-processed as \non{} ($n=95$) were removed.

\begin{table}[h]
  \caption{Multi-class classification performance of the LLM in identifying logical fallacies for the full data ($N=780$) and subset data ($N=685$, where 95 instances of \non{} were removed).}
  \label{tab:perfmetrics}
  \begin{tabular}{ccccc}
    \toprule
    Data & Accuracy & \makecell{Macro\\Precision} & \makecell{Macro\\Recall} & \makecell{Macro\\F1 Score}\\
    \midrule
    Full & 0.79 & 0.76 & 0.66 & 0.71\\
    Subset & 0.90 & 0.87 & 0.89 & 0.88\\
    \bottomrule
  \end{tabular}
\end{table}

The LLM achieved an average accuracy of 0.79 in classifying logical fallacies for the full data.
Many of the inaccuracies occurred in failing to identify a valid logical fallacy, e.g., classifying a text as \non{} (Table~\ref{tab:none}). Most strikingly, the LLM performed worst in identifying instances of the \emo{} fallacy, failing to do so nearly 40\% of the time. As LLMs struggle with extracting fine-grained structured sentiment and opinion information~\cite{zhang2023sentiment}, this may explain the poor performance for \emo{} which demands stronger nuanced interpretations of texts from the LLM.

\begin{table*}[htb!]
  \caption{Breakdown of instances in the full data ($N=780$) that were classified as \non{}.}
  \label{tab:none}
  \begin{tabular}{cccccccc|c}
    \toprule
    & \makecell{Against the\\Person} & \makecell{Appeal to\\Authority} & \makecell{Appeal to\\Popularity} & \makecell{Appeal to\\Emotion} & 
    \makecell{Hasty\\Generalization} & \makecell{Questionable\\Cause} & Red Herring & Total\\
    \midrule
    \non{} & 7 & 4 & 14 & 16 & 27 & 17 & 10 & 95\\
    All & 157 & 74 & 133 & 41 & 165 & 126 & 84 & 780\\
    \midrule
    Percentage & 0.04 & 0.05 & 0.11 & 0.39 & 0.16 & 0.13 & 0.12 & 0.12\\
    \bottomrule
  \end{tabular}
\end{table*}

\begin{figure*}[h]
    \centering
    \includegraphics[width=.8\linewidth]{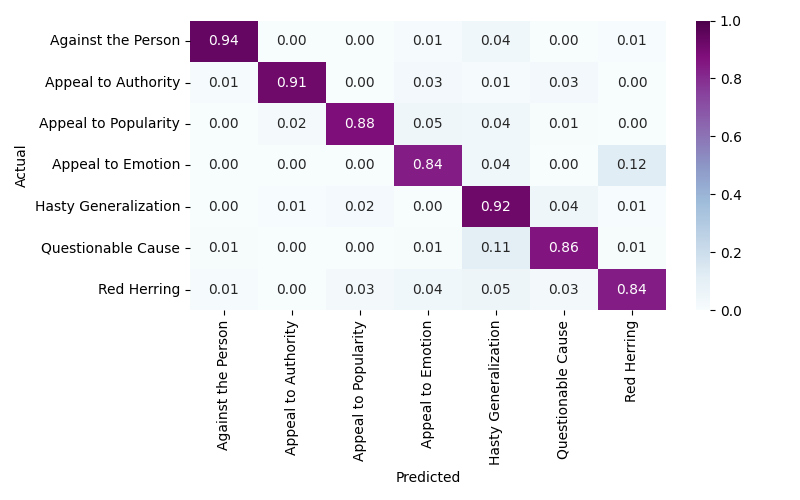}
    \caption{Normalized confusion matrix of the LLM in identifying logical fallacies for the subset data ($N=685$).}
    \label{fig:cmatrix}
\end{figure*}

When excluding the instances classified as \non{}, the LLM achieved an average accuracy of 0.90 on the subset data.
Figure~\ref{fig:cmatrix} shows the normalized classification matrix of the subset data which provides a clearer picture of the misclassifications of one logical fallacy as another. The LLM performed best for \per{} with an accuracy of 0.94 and generally had good performance across the other fallacies, with the lowest accuracy being 0.84. The main misclassifications of logical fallacies occurred for \emo{} where 12\% were classified as \her{} and for \cau{} where 11\% were classified as \gen{}. While the inadequacies of LLMs in making nuanced interpretations may explain the former, the latter could have been due to both fallacies being logically different yet sharing semantically similar definitions (see the definitions in Appendix~\ref{prompt}).

\section{Discussion}

We describe our views of the LLM for our use case of identifying logical fallacies and reflect on the design of the evaluation procedure.

\subsection{Evaluate to Inform Design Decisions}

While the LLM failed to identify valid logical fallacies for more than a tenth of a time (Table~\ref{tab:none}), it performed well in classifying logical fallacies (Figure~\ref{fig:cmatrix}). Since our planned use case prioritizes categorizing the \textit{right} logical fallacy and not surfacing \textit{all} of them, we consider the LLM to be capable and suitable for our task definition.

Without thorough evaluation of the LLM, we could not have had the same level of confidence in choosing to adopt the LLM for our use case. While the results turned out to be in our favor this time, it would not be difficult to imagine how unfavorable results would impact our next steps instead. For instance, if our intervention prioritized finding \textit{all} logical fallacies, we would have made the decision to exclude the \emo{} fallacy going forward due to the subpar performance of the LLM in spotting it.

Such decisions are not naive, underscoring the importance of evaluating LLMs rigorously if they are used as a key component of the research project to inform the design decisions of the artifact or use case where the LLM is adopted.

\subsection{Considerations in Evaluating LLMs}

There were several considerations we faced in the design of the evaluation procedure and we share the more prominent ones here.

\paragraph{\textbf{Evaluation Strategy}}

The first is with choosing the evaluation strategy. Our choice to evaluate the LLM against a labeled dataset was inspired by research from the NLP community where factors like ease of access to existing datasets and convenience in the evaluation process were taken into account. Other evaluation strategies can also be gleaned from the NLP community, which we mention briefly for researchers interested in conducting their own evaluations. A strategy is to engage raters, either humans or automated, to rate the outputs of the LLM according to pre-specified metrics based on theoretical frameworks, typically using Likert scales. Another strategy is to find existing benchmarks that are applicable to the intended use case and to run the outputs of the LLM against them.

\paragraph{\textbf{Dataset}}

The second is with finding the appropriate dataset. The dataset we used contained mostly instances with one or two sentences and occasionally those with a paragraph. Yet, for our planned use of the LLM, we intend to pass online discussion comments that can range from a few sentences to a few paragraphs. While our evaluation shows that the LLM performs acceptably for shorter texts, the same may not be expected for longer texts which are likely to contain more logical fallacies. This would merit further testing, although to the best of our knowledge, there is no existing labeled dataset on logical fallacies for long texts which would necessitate manually creating one. Nonetheless, we will acknowledge this as a limitation of the evaluation of the LLM when we report about its adoption in the digital misinformation intervention.

\paragraph{\textbf{Prompt Engineering}}

The last is with settling on the prompt. While not reported here, several iterations of the prompt were made before the final version. While prompts can be easily created, what is more difficult is deciding when to \textit{stop} engineering. While we initially focused on crafting a prompt that would achieve the best performance, we reflected on our priorities as HCI researchers and felt that more so than the best performance, what is important is to design the interactions of the artifact with an awareness of the capabilities and insufficiencies of the LLM for the task. Refocusing, we aimed for a prompt that would deliver adequate performance, as presented in this work, and to gain a comprehensive understanding of the extent of its performance. Here, we also want to highlight how designing and evaluating LLMs prompts became an iterative process for us where an established evaluation protocol lended towards knowing when to call a halt.

\section{Conclusion}

While the use of LLMs for HCI research is gaining interest, the question of whether LLMs are suitable for their intended tasks remains. HCI researchers should go beyond incorporating the LLM merely because it ``seems to work'' and instead conduct a thorough evaluation for its specific use. In line with this, we present an evaluation of the use of an LLM in identifying logical fallacies, describing our approach, results and reflections. Keeping in mind that the LLM performs adequately, though not perfectly, for this task, we intend to proceed with using the LLM for our research, being transparent about its performance with the users.

\begin{acks}
This work was supported by resources from the KAIST Interaction Lab (KIXLAB).
\end{acks}

\bibliographystyle{ACM-Reference-Format}
\bibliography{main}

\newpage

\appendix

\section{Prompt for Identifying Logical Fallacies}\label{prompt}

\begin{lstlisting}
There are seven fallacies. Each fallacy, its definition, and examples of texts with the fallacy are provided below.

Fallacy: hasty generalization
Definition: Drawing a conclusion about all or many instances of a phenomenon on the basis of one or a few instances of that phenomenon.
Examples:
1) "Four out of five dentists recommend Happy Glossy toothpaste. Therefore, it must be great."
2) No one in your family surfs? But I thought you said you lived in California before this.
3) All chess players are geniuses.

Fallacy: questionable cause
Definition: Concluding that one thing caused another simply because they are regularly associated.
Examples:
1) Children who play violent video games act more violently than those who don't.
2) "President Kumail raised taxes, and then the rate of violent crime went up. Kumail is responsible for the rise in crime."
3) People who eat yogurt have healthy guts. If I eat yogurt I will never get sick.

Fallacy: ad populum
Definition: Affirming that something is real or better because the majority in general or of a particular group thinks so.
Examples:
1) Everyone seems to support the changes in the vacation policy, and if everyone likes them, they must be good.
2) "Since 88% of people polled believe in UFOs, they must exist."
3) Don't be the only one not wearing Nike!

Fallacy: ad hominem
Definition: Attacking the person or some aspect of the person making the argument instead of addressing the argument directly.
Examples:
1) Only a selfish, non-caring person would believe that this is ok.
2) You can't believe Jack when he says there is a God because he doesn't even have a job.
3) The reason our company never makes any money is because we have a buffoon running it!

Fallacy: appeal to emotion
Definition: Manipulating the reader's emotions in order to win the argument in place of a valid reason.
Examples:
1) Regime vs. government Pro-death vs. pro-choice
2) "It is an outrage that the school wants to remove the vending machines. This is taking our freedom away!"
3) Television Advertisement: Get all of your stains out by using new and improved Ultra Suds and wash your blues away!

Fallacy: red herring
Definition: Attempting to divert the reader's attention from the original argument by offering a different point.
Examples:
1) I shouldn't be punished for staying out past my curfew because I did the dishes earlier today.
2) "Yes, we have safety issues in our factory. But we work really hard to make a good product. They sell so well!"
3) Mining may destroy the environment, but how about the people whose jobs will be affected? Should we ban it?

Fallacy: appeal to authority
Definition: Using an alleged authority who is not really an authority on the facts relevant to the argument as evidence.
Examples:
1) My professor, who has a Ph.D. in Astronomy, once told me that ghosts are real. Therefore, ghosts are real.
2) There is definitely a link between dementia and drinking energy drinks. I read about it on Wikipedia.
3) Coke is not as healthy for you as Pepsi. Besides, Britney Spears drinks Pepsi, so it must be healthier than Coke.

For the text below, identify if there are any fallacies {hasty generalization, questionable cause, ad populum, ad hominem, appeal to emotion, red herring, appeal to authority}.

For each part of the text with a fallacy, use this template.
Part: {quote verbatim the part of the text with the fallacy}
Fallacy: {fallacy}

If the text does not contain any fallacy, use this template.
Part: none
Fallacy: none

The text:
--The text goes here--
\end{lstlisting}

\end{document}